\documentclass[%
reprint,
 amsmath,amssymb,
 aps,
prb,
floatfix
]{revtex4-2}

\usepackage{graphicx}
\usepackage{dcolumn}
\usepackage{bm}

\usepackage{braket} 

\begin{document}


\title{Semiconductor nanofilms as thermal phonon polarizers: competing effects of scattering selection rules and boundary mode conversion}

\author{Vasumathy Ravishankar}
\author{Navaneetha K. Ravichandran}
\email{navaneeth@iisc.ac.in}
\affiliation{%
 Department of Mechanical Engineering, Indian Institute of Science, Bangalore 560012, India
 }%

\date{\today}

\begin{abstract}
Phonon scattering selection rules are known to control heat flow through bulk solids. Here we show that these selection rules also modulate heat flow through nanoscale semiconductor films, although through a previously-unexplored mechanism. Using first-principles calculations, we expose a competition between these selection rules and phonon mode conversion at boundaries of nanoscale films, that drives mode-polarized heat currents at cryogenic temperatures ($\le$ 100 K). This polarizing effect is stronger in materials like indium phosphide, where selection rules based on large velocity differences between phonon branches amplifies the longitudinal acoustic (LA) phonon contribution to thermal conductivity by restricting their intrinsic scattering events, while boundary mode conversion in nanoscale films suppresses it by depopulating the LA phonons. The resulting transverse-polarized non-equilibrium phonons will enable symmetry-selective engineering of phonon coupling to electrons, strains and defects in nanoscale films, that is difficult to achieve in bulk solids.
\end{abstract}

\maketitle
\clearpage
Selection rules for phonon scattering~\cite{lindsay_first-principles_2013, ravichandran_phonon-phonon_2020}, which originate from momentum and energy conservation requirements during the scattering events, have played a crucial role in the discovery of materials with intriguing thermal properties. The most well-known of these is the discovery of a new ultrahigh thermal conductivity ($\kappa$) material that is not present in nature – boron arsenide (BAs), whose unusual phonon dispersion activates two scattering selection rules that sharply suppresses the lowest-order scattering events involving three phonons, thus culminating in the first set of experiments to confirm the strong role of higher-order four-phonon scattering in affecting the $\kappa$ of bulk BAs crystals~\cite{tian_unusual_2018, li_high_2018, kang_experimental_2018}. These selection rules have also been instrumental in driving unconventional pressure dependencies of the ultrahigh-$\kappa$’s of BAs~\cite{ravichandran_non-monotonic_2019, li_anomalous_2022}, boron phosphide (BP) and silicon carbide (SiC)~\cite{ravichandran_exposing_2021}, an unusually strong isotope-effect on the $\kappa$ of BP~\cite{zhu_vapor-flux_nodate} and the ultrahigh $\kappa$ of semi-metallic $\theta$-phase tantalum nitride~\cite{kundu_ultrahigh_2021, li_metallic_2026}. Thus, these selection rules have fundamentally upended our understanding of thermal phonon scattering in bulk semiconductors, which were originally based on simple empirical observations from several decades ago~\cite{slack_thermal_1979}.

While the influence of these selection rules on the $\kappa$’s of bulk crystals is well established, we demonstrate, using first-principles calculations, that they impact the $\kappa$’s of thin semiconductor films as well, although through a previously-unrecognized mechanism. We show that these selection rules compete with mode conversion events that occur when phonons reflect off the boundaries of thin films, to produce phonon mode-polarized heat currents, particularly at cryogenic temperatures (T $\le$ 100 K). We observe the strongest manifestation of this effect in thin films of indium compounds with group V elements, where the scattering selection rule originating from the large differences in the group velocities of the longitudinal (LA) and the transverse (TA) acoustic phonons sharply weaken the intrinsic scattering rates of the LA phonons, thus amplifying their contribution to $\kappa$, while the boundary mode conversion rapidly depopulates them, resulting in a TA-polarized heat current with significantly lower $\kappa$ compared to the bulk. Since phonon coupling to strain fields, defects and electrons is polarization-selective, our prediction of TA-polarized non-equilibrium phonons in nanoscale semiconductor films will drive symmetry-informed tuning of $\kappa$'s and hot-carrier lifetimes in these films, that is challenging in bulk crystals.

We predict the effect of phonon mode conversion at thin film boundaries on the non-equilibrium thermal phonons using a first-principles variance-reduced Monte Carlo (VRMC) solution of the steady-state Peierls-Boltzmann equation (SSPBE) for phonon transport~\cite{peraud_efficient_2011, peraud_alternative_2012, ravichandran_coherent_2014}. For the materials considered here, it is sufficient to solve the SSPBE under the relaxation time approximation (RTA), given by:
\begin{align}
    \mathbf{v}_{\lambda}\cdot\nabla_{\textbf{r}} e_\lambda = -\frac{e_\lambda - e^0_\lambda\left(T\right)}{\tau_\lambda}  \label{eq:SSPBE}
\end{align}
where $e_\lambda$ is the unknown non-equilibrium phonon energy distribution function for a mode $\lambda\equiv\left(\nu, p\right)$ of frequency $\nu$ and polarization $p$, $e^0_\lambda\left(T\right)$ is the corresponding equilibrium energy distribution at a local temperature $T$ and $\tau_\lambda$ is the phonon relaxation time calculated using the Matthiessen's rule with contributions from three-phonon, four-phonon and phonon-isotope interactions obtained from our previous work~\cite{ravichandran_phonon-phonon_2020}. VRMC simulation is a stochastic method of solving the SSPBE [Eq.~\ref{eq:SSPBE}], where computational particles representing phonon bundles are initialized, advected and scattered according to the SSPBE directly in the thin films, and sampled in space to obtain the heat flux, the temperature distribution and hence, the thin film $\kappa$. We generate the initial distribution of phonons volumetrically by drawing samples from the distribution $\sum_\lambda D_\lambda e^d_\lambda v^{\parallel}_\lambda$~\cite{peraud_alternative_2012}, where $D_\lambda$ is the phonon density of states (DOS), $e^d_\lambda = e_\lambda - e^0_\lambda\left(T_0\right)$, $T_0$ is the global equilibrium temperature and $v^{\parallel}_\lambda$ is the group velocity component along the negative temperature gradient, $\left[-\frac{\mathrm{d}T}{\mathrm{d}x}\right]$, parallel to the thin film boundary. The initial direction of propagation is chosen such that its component along $\left[-\frac{\mathrm{d}T}{\mathrm{d}x}\right]$ is uniformly distributed on a hemisphere to support a steady-state heat flux, as described in Ref.~\cite{peraud_efficient_2011}. During intrinsic scattering events, phonons are deleted and new phonons are drawn from $\sum_\lambda \frac{D_\lambda e^d_\lambda}{\tau_\lambda}$, thus mimicking the collision operator of the SSPBE [Eq.~\ref{eq:SSPBE}]. Energy conservation is implicitly satisfied by enforcing the same effective energy for all computational particles~\cite{peraud_alternative_2012}. 

To solve the SSPBE in the $\nu$-formulation, the wave vector-dependent group velocities and relaxation times are projected into an isotropic form as described in Ref.~\cite{hua_analytical_2014} - an approximation that has been successfully used to predict heat flow in thin films of cubic crystals~\cite{minnich_determining_2012, ravichandran_spectrally_2018, jeong_transient_2021}, such as those in this work. We account for boundary mode conversion by statistically conserving the incoming and outgoing heat fluxes normal to the adiabatic boundaries of the films, i.e.,
\begin{align}
\sum_p^{\left(\text{in}\right)} D_\lambda e^d_\lambda v^{\perp}_\lambda = \sum_p^{\left(\text{out}\right)} D_\lambda e^d_\lambda v^{\perp}_\lambda	\label{eq:mode_conv_flux}
\end{align}
where $v^{\perp}_\lambda$ is the component of the phonon group velocity perpendicular to the thin film boundary, with the sum being only over the polarizations $p$ at a fixed frequency $\nu$ to represent elastic scattering. Enforcing Eq.~\ref{eq:mode_conv_flux} in the VRMC solution amounts to deleting the incoming phonons at the boundary and re-emitting phonons drawn from $\sum_p^{\left(\text{out}\right)} D_\lambda e^d_\lambda v^{\perp}_\lambda$ for the same $\nu$. We determine the angle of reflection for specular scattering from Snell's law, which preserves the component of phonon momentum ($\hbar \vec{q}$, where $\vec{q}$ is the wave vector) along the boundary upon reflection with mode conversion. For diffuse reflections, we choose the angle of emission uniformly on a hemisphere facing into the film at the boundary~\cite{peraud_efficient_2011}, regardless of whether mode conversion is allowed.

\begin{figure}[!ht]
\centering
\includegraphics[width=\linewidth, trim=12mm 12mm 0mm 10mm, clip]{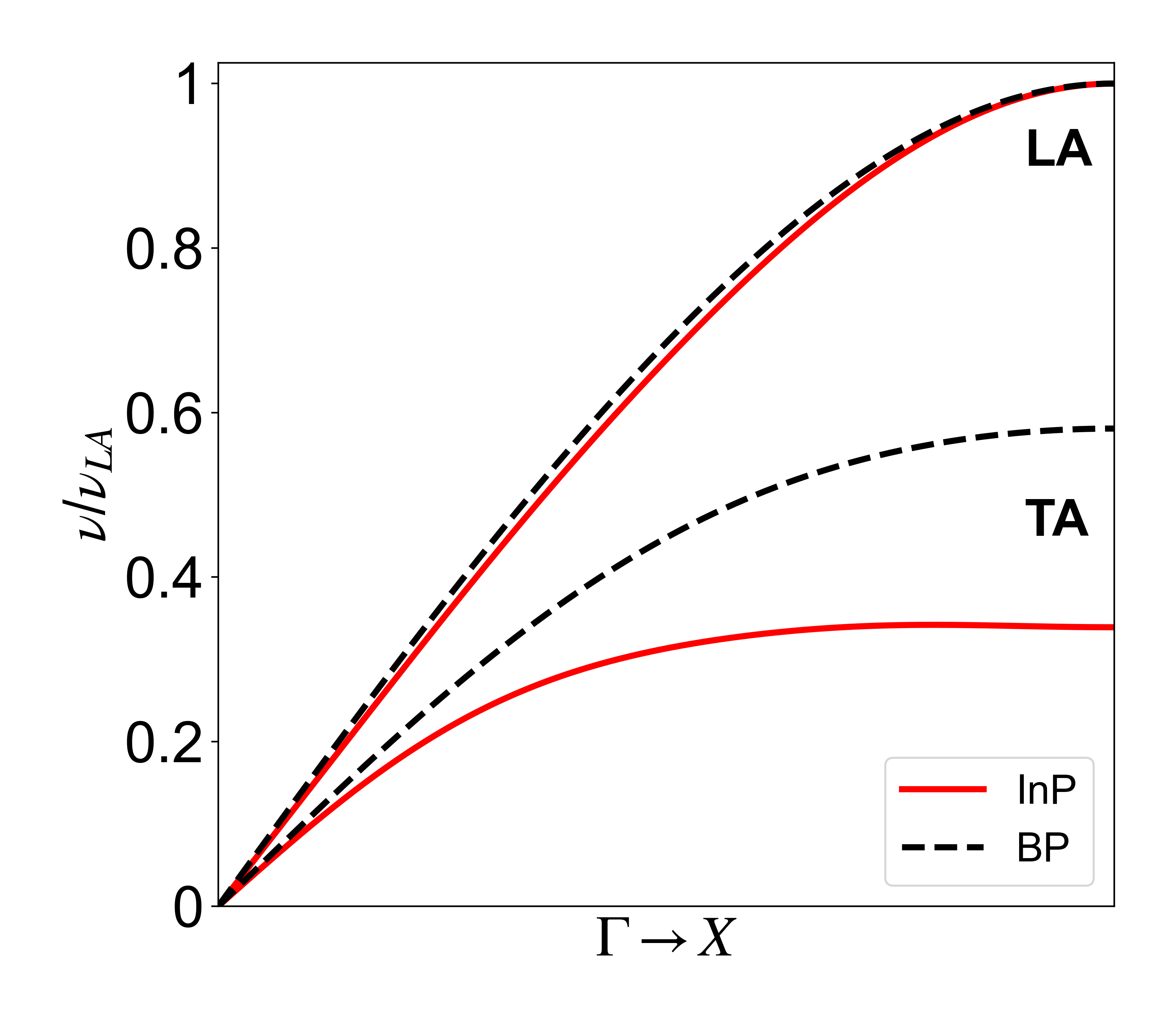}
\caption{Acoustic phonon dispersions of InP and BP along the $\Gamma\to X$ high symmetry direction. The phonon dispersions along other high-symmetry directions for these materials are shown in Supplementary Figs. S1(a)-(b). LA and TA phonon dispersions are farther apart in InP than in BP, thus activating the AAA \# 2 selection rule in InP~\cite{ravichandran_phonon-phonon_2020}.}  \label{fig:Figure1}
\end{figure}

\begin{figure*}[!ht]
\centering
\includegraphics[width=\linewidth, trim=4mm 10mm 10mm 5mm, clip]{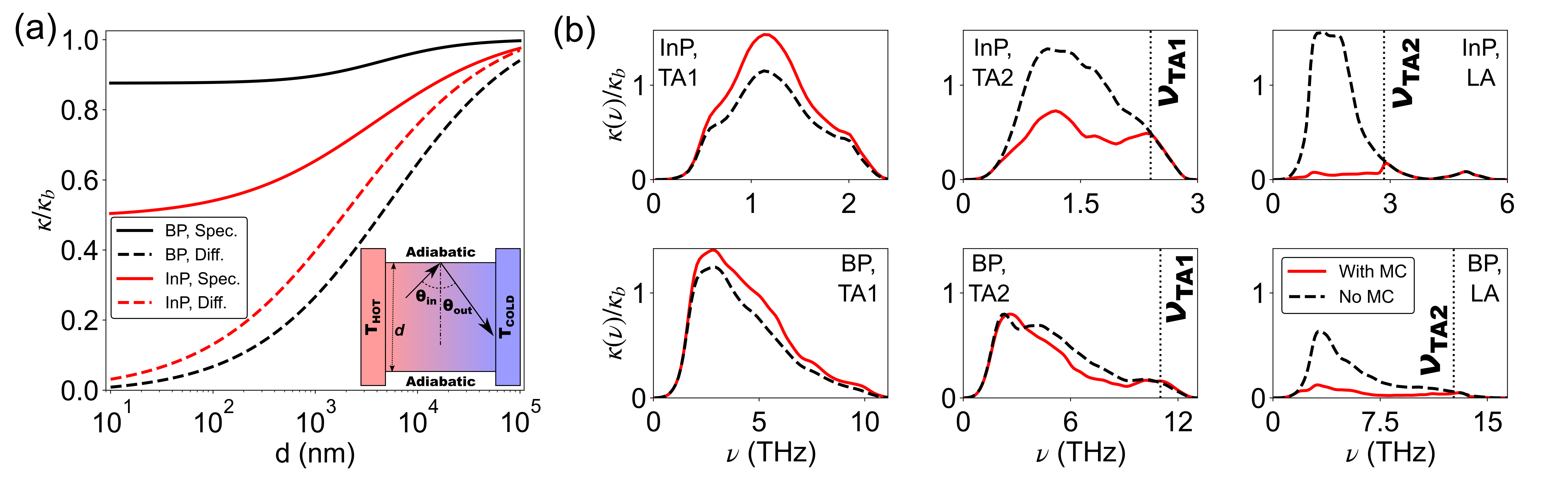}
\caption{Effect of mode conversion on the $\kappa$ of BP and InP thin films at 100 K. (a) Thin film $\kappa$ of BP and InP vs. thickness ($d$) normalized by the corresponding bulk values ($\kappa_b$) at 100 K. The inset shows the simulation geometry, and the incidence ($\theta_{\text{in}}$) and the reflection ($\theta_{\text{out}}$) angles for a phonon undergoing mode conversion at the thin film boundary. For specular reflection without mode conversion, the $\kappa$'s of thin films are equal to $\kappa_b$, independent of $d$~\cite{ravichandran_role_2016}. When mode conversion is allowed, the predicted $\kappa$'s for specular reflections decrease faster with $d$ in InP than in BP. For diffuse reflections, the predicted $\kappa\left(d\right)$'s remain the same with or without mode conversion in both materials, hence not shown separately. (b) Spectral contributions to $\kappa$ normalized by the corresponding bulk value ($\kappa_b$) from the three acoustic branches in thin films of InP and BP with $d$ = 100 nm at 100 K, for the specular boundary reflection with (With MC) and without (No MC) mode conversion. The area under each curve represents the contribution of each phonon branch to the overall $\kappa$ normalized by the $\kappa_b$ of the corresponding material. $\nu_{\text{TA1}}$ and $\nu_{\text{TA2}}$ are the maximum frequencies of the two TA branches. Scattering selection rules amplify the contribution of the LA phonons to $\kappa$ in bulk InP, which is also observed in InP nanofilms when specular boundary reflections of phonons occur without mode conversion, but is strongly suppressed when mode conversion is permitted. The effect of this interplay on the $\kappa$ of thin BP films is minimal, since the TA phonons are the dominant heat carriers in BP.}    \label{fig:figure2}
\end{figure*}

To elucidate our findings, we focus on two representative materials - naturally-occurring forms of BP and indium phosphide (InP), with contrasting phonon dispersions, as shown in Fig.~\ref{fig:Figure1}. In BP, the acoustic phonons are bunched together, which suppresses their scattering rates within the framework of three-phonon scattering via the AAA \# 1 selection rule [AAA $\equiv$ all-acoustic three-phonon scattering] described in Ref.~\cite{ravichandran_phonon-phonon_2020, ravichandran_exposing_2021}. These weak scattering rates among the acoustic phonons are only partially exposed at 100 K due to phonon-isotope scattering [Supplementary Fig. S2(a)], thus preferentially enhancing the contribution of low-frequency TA phonons to overall $\kappa$ at 100 K in bulk natural BP crystals, as shown in Fig.~\ref{fig:figure2} (b). In stark contrast, the LA phonon dispersion of natural InP is much steeper than those of the TA phonons, resulting in the activation of the AAA \# 2 selection rule described in Ref.~\cite{ravichandran_phonon-phonon_2020} that preferentially weakens the AAA scattering rates of low frequency LA phonons [Supplementary Fig. S2(b)], thus amplifying their contribution to overall $\kappa$ in bulk InP crystals at 100 K [Fig.~\ref{fig:figure2} (b)]. For natural BP and InP, the first-principles bulk $\kappa$ values ($\kappa_b$) at 100 K from the SSPBE under the RTA (full iterative solution) are 2589 Wm$^{-1}$K$^{-1}$ (2906 Wm$^{-1}$K$^{-1}$) and 337 Wm$^{-1}$K$^{-1}$ (347 Wm$^{-1}$K$^{-1}$) respectively; hence, the RTA is sufficient for these materials at 100 K.

The calculated $\kappa$'s of BP and InP thin films at 100 K shown in Fig.~\ref{fig:figure2} (a) exhibit two important features. First, the calculated suppression in $\kappa$ due to diffuse boundary scattering agree with the predictions from the Fuchs-Sondheimer theory~\cite{ravichandran_role_2016}, regardless of whether mode conversion occurs at the boundaries. Second, for specular boundary reflections with mode conversion, the $\kappa$'s are suppressed compared to the corresponding $\kappa_b$'s for both BP and InP, with stronger suppression for InP films than BP films. On the other hand, specular reflections without mode conversion do not offer any additional resistances to heat flow in thin films~\cite{ravichandran_role_2016, ravichandran_spectrally_2018}.

The observations for diffuse scattering can be understood by tracking the change in the individual phonon contribution to $\kappa$ after every scattering event in the VRMC simulation. When a phonon undergoes diffuse boundary reflection, the direction of propagation is randomized uniformly, regardless of whether mode conversion occurs. As discussed in Ref.~\cite{peraud_alternative_2012}, the phonon contribution to $\kappa$ drops by two orders of magnitude after any such scattering event that randomizes its propagation direction. Therefore, for diffuse reflections, the phonon contribution to $\kappa$ remains significant only until its first intrinsic or boundary scattering event, which remains unaffected by the possibility of mode conversion during the impending boundary reflection event.

\begin{figure}[!ht]
\centering
\includegraphics[width=\linewidth, trim=5mm 12mm 5mm 10mm, clip]{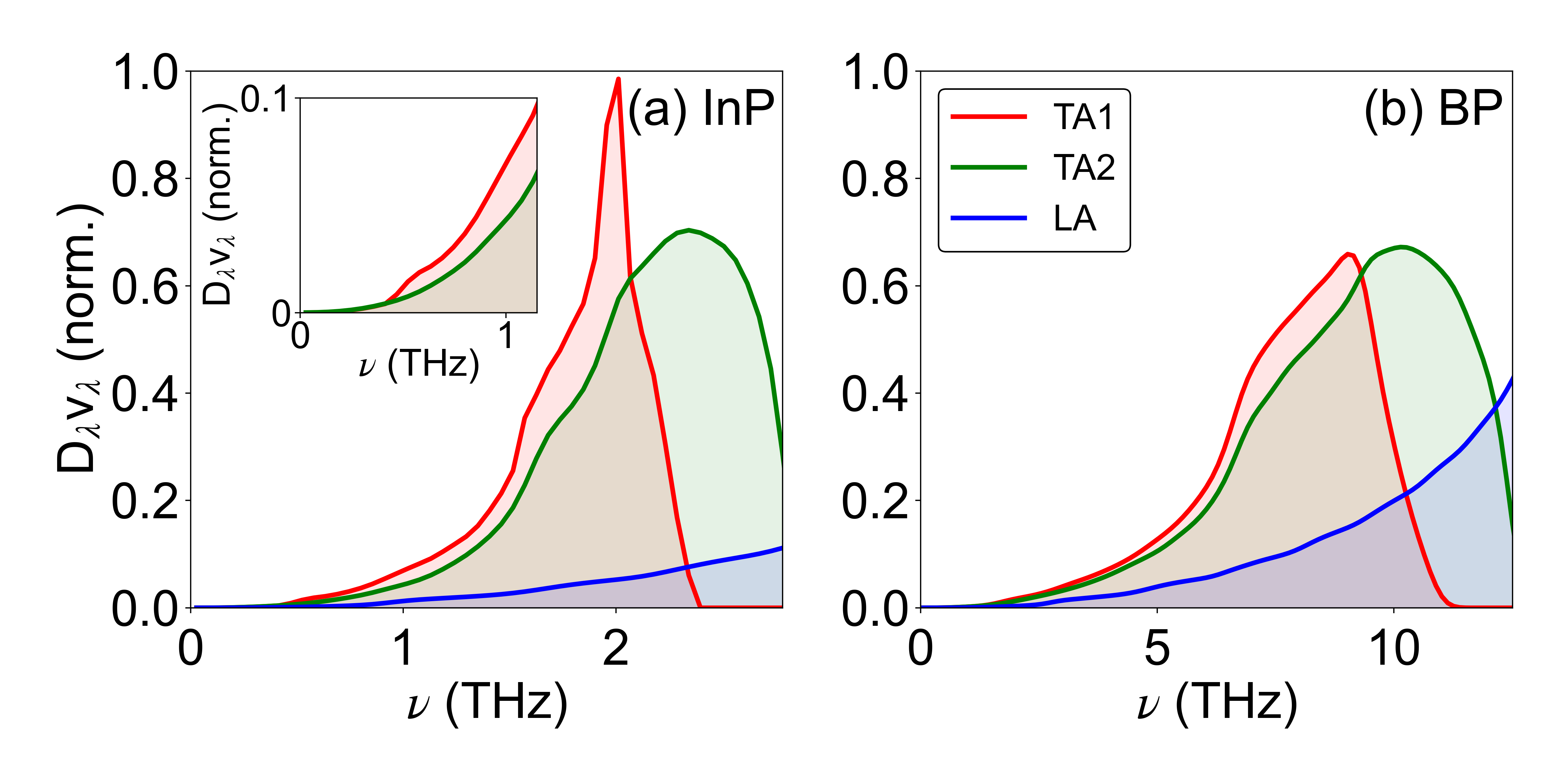}
\caption{The normalized material factor $D_\lambda v_\lambda$ of the mode conversion probabilities in InP (a) and BP (b). The x-axis is truncated at the maximum $\nu$ of the TA$_2$ branch, beyond which mode conversion does not occur for the LA phonons [Fig.~\ref{fig:figure2} (b)], since $D^{\text{TA1}}_\lambda = D^{\text{TA2}}_\lambda = 0$. In both materials, $D_\lambda v_\lambda$ is much larger for the TA phonons than for the LA phonons, with the contrast being greater for InP than BP, due to the large velocity differences of the phonon polarizations in the former.}  \label{fig:Figure3}
\end{figure}

For specular reflections with mode conversion, the thin film $\kappa$ progressively deviates from $\kappa_b$ as the film thickness ($d$) decreases, with the observed deviation being material-dependent [Fig.~\ref{fig:figure2} (a)]. It saturates at about 88\% of $\kappa_b$ at $d\sim 500$ nm for BP, while continuing to decrease till $d \sim$ 10 nm, reaching down to $\sim$ 50\% of $\kappa_b$ for InP at 100 K. The stronger $\kappa$-suppression in InP nanofilms originates from the sharp reduction in the $\kappa$-contribution of the LA phonons due to mode conversion, as shown in Fig.~\ref{fig:figure2} (b). When a phonon with polarization $p$ reflects off a film boundary, it undergoes mode conversion into other polarizations ($p' \ne p$) or remains in the same polarization ($p' = p$) with relative probabilities $\left[D\left(\nu, p'\right)v^{\perp}\left(\nu, p'\right)\right]_{\text{out}}$, according to Eq.~\ref{eq:mode_conv_flux}. These relative probabilities contain two parts: the factor $D\left(\nu, p\right)v\left(\nu, p\right)\equiv D_\lambda v_\lambda$ that depends on the phonon properties of the bulk material only, and the geometric factor $\cos\theta_{\text{out}}$ that projects the velocity of the reflected phonon normal to the boundary [see schematic in Fig.~\ref{fig:figure2} (a)]. Figure~\ref{fig:Figure3} shows that the bulk material factor, $D_\lambda v_\lambda$, is much larger for the TA phonons than for the LA phonons in both materials. Furthermore, InP shows a greater contrast in $D_\lambda v_\lambda$ between the LA and the TA branches compared to BP, since the larger separation between the LA and the TA branches in InP [Fig.~\ref{fig:Figure1}] causes a larger difference in their $D_\lambda$ and $v_\lambda$. Hence, an incoming LA phonon has a high probability of converting into one of the TA polarizations, while the reciprocal TA $\to$ LA conversion is unlikely, with larger disparities in InP than in BP. 

Apart from the material factor, the angle of specular boundary reflection, $\theta_{\text{out}}$, that enters into the geometric factor, $\cos\theta_{\text{out}}$, also acts to TA-polarize the heat current in thin films. For every reflected phonon, $\theta_{\text{out}}$ is determined using Snell's law to conserve the phonon momentum parallel to the boundary as $\sin\theta_{\text{out}} = \frac{q\left(\nu, p\right)}{q\left(\nu, p'\right)}\sin\theta_{\text{in}}$, where $\theta_{\text{in}}$ is the angle of incidence, $p$ and $p'$ are the incoming and the outgoing phonon polarizations respectively and $q\left(\nu, p\right)$ is the magnitude of the wave vector. Figure~\ref{fig:Figure1} shows that $\left[\frac{q\left(\nu, \text{TA}\right)}{q\left(\nu, \text{LA}\right)}\right]^{\text{InP}} > \left[\frac{q\left(\nu, \text{TA}\right)}{q\left(\nu, \text{LA}\right)}\right]^{\text{BP}} > 1$ at a fixed $\nu$. Thus, while the Snell's law allows all incident LA phonons to convert into TA modes, the requirement of $\sin\theta_{\text{out}} < 1$ restricts the reciprocal TA $\to$ LA mode conversion to only those incoming TA phonons with $\theta_{\text{in}}$ less than a critical angle $\theta_{\text{c}} = \sin^{-1}\left[\frac{q\left(\nu, \text{LA}\right)}{q\left(\nu, \text{TA}\right)}\right] < 90^o$, with $\theta_{\text{c}}^{\text{InP}} < \theta_{\text{c}}^{\text{BP}}$ due to the larger velocity differences of phonon polarizations in InP. Any possible evanescent and surface LA fields that could emerge for $\theta_{\text{in}} > \theta_c$ due to mode conversion from an incoming TA phonon are localized to $\sim$nm-thick regions near the boundary for THz-frequency thermal phonons, and so, do not contribute significantly to the heat current. Therefore, the geometric effect from Snell's law also preferentially depopulates the LA modes, thus further aiding in the generation of TA-polarized non-equilibrium phonons, with stronger polarization in InP than in BP.
\begin{figure}[!ht]
\centering
\includegraphics[width=\linewidth, trim=10mm 12mm 10mm 10mm, clip]{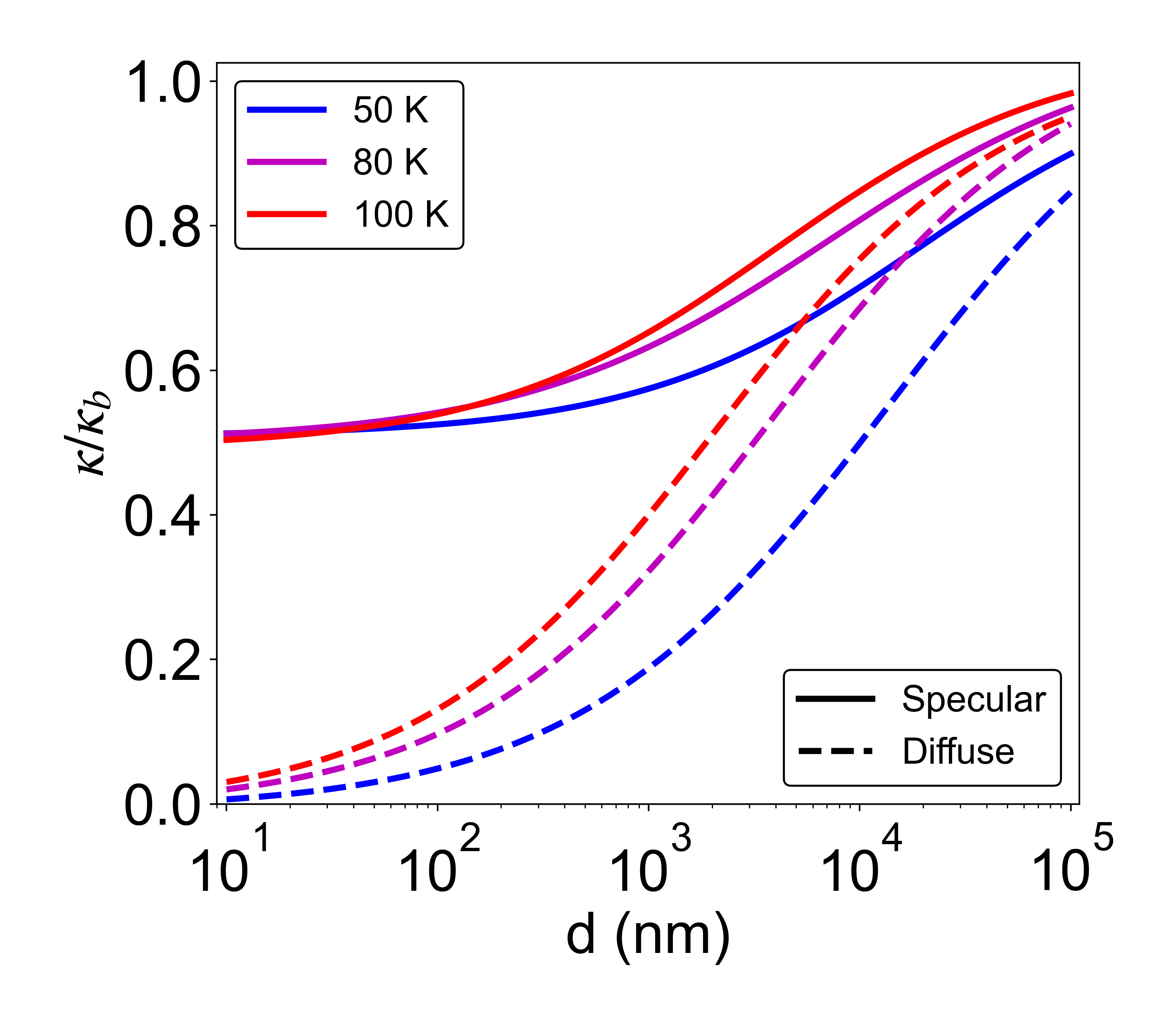}
\caption{Thickness-dependent $\kappa$ of InP films at different cryogenic temperatures scaled by $\kappa_b$. When phonons undergo specular reflections with mode conversion, $\kappa/\kappa_b$ saturates at larger
$d$ as $T$ is lowered, due to the longer phonon MFPs at lower $T$. For diffuse reflections, mode conversion has no effect on $\kappa\left(d\right)/\kappa_b$.}  \label{fig:Figure4}
\end{figure}

Finally, the observed saturation of $\kappa/\kappa_b$ at small $d$ in Fig.~\ref{fig:figure2} (a) originates from size effects in these nanoscale films. As described earlier, the component of the initialized phonon velocity along $\left[-\frac{\mathrm{d}T}{\mathrm{d}x}\right]$ is uniformly distributed on a hemisphere in the VRMC simulation. Thus, with decreasing $d$, boundary reflection with mode conversion precedes intrinsic scattering for increasingly large number of initialized phonons, since the likelihood of their trajectories intersecting with the film boundaries before any intrinsic scattering event increases. At sufficiently small $d$, the entire initialized phonon population undergoes boundary mode conversion first before intrinsic scattering, thereby attaining the distribution in the right hand side of Eq.~\ref{eq:mode_conv_flux} without any memory of the initialized distribution. Since this phonon distribution from mode conversion originates from Eq.~\ref{eq:mode_conv_flux}, it is unaffected by any subsequent boundary mode conversion events. Any further reduction in $d$ will not modify this distribution, since Eq.~\ref{eq:mode_conv_flux} is independent of $d$. $\kappa/\kappa_b$ saturates at larger values and at larger $d$ in BP films compared to InP films in Fig.~\ref{fig:figure2} (a) due to the smaller LA phonon contribution to $\kappa_b$ and longer intrinsic phonon mean free paths (MFP) [Supplementary Fig. S1(c)] in the former. 

As a consequence of this size effect, we find from  Fig.~\ref{fig:Figure4} that the saturation of $\kappa/\kappa_b$ in InP films occurs at larger $d$ as $T$ is lowered, due to the longer phonon MFPs at lower $T$ caused by the reduced thermal occupation of the phonons that participate in scattering, with $\kappa/\kappa_b$ saturating even for films as thick as 500 nm at 50 K. Interestingly, $\kappa/\kappa_b$ saturates to the same value at small $d$ in this range of $T$ due to mode conversion in Fig.~\ref{fig:Figure4}. We find that this $T$-independent small-$d$ limit of $\kappa/\kappa_b$ originates from $T$-independent relative contributions of the LA and TA phonons to $\kappa_b$ between 50 K and 100 K ($\sim$ 40\% for LA and $\sim$ 60\% for TA). Our calculations show that similar trends of $\kappa$ occur in thin films of other indium compounds with AAA \# 2 selection rule activated [indium arsenide (InAs) and indium antimonide (InSb)], although the effect is weaker than in InP films [Supplementary Fig. S3] due to the smaller frequency gap between the acoustic and the optic phonons in InAs and InSb compared to InP, thus resulting in a partial masking of the effect of the AAA \# 2 selection rule by the scattering among two acoustic and one optic phonons [AAO channel], as previously discussed in Ref.~\cite{ravichandran_phonon-phonon_2020}.

In summary, we demonstrate, using first-principles calculations, that the large contribution of the LA phonons to $\kappa$, driven by the activation of the AAA \# 2 scattering selection rule in bulk InP crystals, is strongly suppressed in nanoscale InP films by the preferential depopulation of the LA phonons due to mode conversion upon specular reflection at film boundaries. This competition between intrinsic scattering selection rules and extrinsic boundary mode conversion suppresses the $\kappa$ of $10-100$ nm thick InP films to $\sim 50\%$ of $\kappa_b$ at 100 K and results in a heat current that is strongly polarized with out-of-equilibrium TA modes. Interestingly, both of these competing effects originate from the large differences in the group velocities of the phonon polarizations in InP. Although the $\kappa$ of thin films are suppressed relative to $\kappa_b$ in films of other materials like BP as well, the effect is weaker, since the LA phonons are not the dominant heat carriers in these materials. Mode-polarized heat currents offer a route to tune hot phonon interactions with electrons, defects and strain fields, since these interactions are polarization-selective. With the recent experimental reports of strong specular reflections of thermal phonons at atomically rough surfaces~\cite{gelda_specularity_2018, ravichandran_spectrally_2018} and the recent advances in fabricating InP films with thicknesses $< 10\ \mu$m~\cite{park_layer-resolved_2022}, our work will inspire applications that benefit from mode-polarized heat currents such as efficient and sensitive electronics with extended hot carrier lifetimes and environmentally-robust devices with defect-insensitive thermal properties for quantum and deep space applications.

This work was supported by the Core Research Grant (CRG) No. CRG/2022/009160, and the Mathematical Research Impact Centric Support (MATRICS) Grant No. MTR/2022/001043 from the Science and Engineering Research Board, India, the Advanced Research Grant (ARG) No. ANRF/ARG/2025/007160/ENS from the Anusandhan National Research Foundation, India, the IISc-IoE post-doctoral fellowship (VR) and the Infosys Young Investigator award (NR).
\bibliographystyle{unsrt}
\bibliography{references_PRL}
\end{document}